\documentclass[aps,twocolumn,showpacs]{revtex4}
\usepackage{amsmath}
\usepackage{graphicx}
\usepackage{dcolumn}
\usepackage{verbatim}
\usepackage{times}
\usepackage{subfigure}
\usepackage{bm}
\usepackage{color}
\usepackage[colorlinks,dvipdfm]{hyperref}
\usepackage{subfigure}

\begin{document}

\title{Phase evolution of the two-dimensional Kondo lattice model near
half-filling}
\author{Huan Li$^{1}$, Yu Liu$^{2,6}$, Guang-Ming Zhang$^{3,5}$, and Lu Yu$%
^{4,5}$}
\affiliation{$^{1}$College of Science, Guilin University of Technology, Guilin 541004,
China \\
$^{2}$ICP, Institute of Applied Physics and Computational Mathematics, Beijing
100088, China \\
$^{3}$State Key Laboratory of Low-Dimensional Quantum Physics and Department
of Physics, Tsinghua University, Beijing 100084, China\\
$^{4}$Institute of Physics, Chinese Academy of Sciences, Beijing 100190,
China\\
$^5$Collaborative Innovation Center of Quantum Matter, Beijing, China\\
$^6$Software Center for High Performance Numerical Simulation, China Academy of Engineering Physics, Beijing 100088, China
}
\date{\today}

\begin{abstract}
Within a mean-field approximation, the ground state and finite temperature
phase diagrams of the two-dimensional Kondo lattice model have been
carefully studied as functions of the Kondo coupling $J$ and the conduction
electron concentration $n_{c}$. In addition to the conventional
hybridization between local moments and itinerant electrons, a staggered
hybridization is proposed to characterize the interplay between the
antiferromagnetism and the Kondo screening effect. As a result, a heavy
fermion antiferromagnetic phase is obtained and separated from the pure
antiferromagnetic ordered phase by a first-order Lifshitz phase transition,
while a continuous phase transition exists between the heavy fermion
antiferromagnetic phase and the Kondo paramagnetic phase. We have developed
a efficient theory to calculate these phase boundaries. As $n_{c}$ decreases
from the half-filling, the region of the heavy fermion antiferromagnetic
phase shrinks and finally disappears at a critical point $n_{c}^{\ast
}=0.8228$, leaving a first-order critical line between the pure
antiferromagnetic phase and the Kondo paramagnetic phase for $%
n_{c}<n_{c}^{\ast }$. At half-filling limit, a finite temperature phase
diagram is also determined on the Kondo coupling and temperature ($J$-$T$)
plane. Notably, as the temperature is increased, the region of the heavy
fermion antiferromagnetic phase is reduced continuously, and finally
converges to a single point, together with the pure antiferromagnetic phase
and the Kondo paramagnetic phase. The phase diagrams with such triple
point may account for the observed phase transitions in related heavy
fermion materials.
\end{abstract}

\maketitle

\section{Introduction}

Since the discovery of heavy-fermion materials, the underlying mechanism
controlling these rare earth or actinide-based compounds has continuously
been the focuses of exploration \cite{Stewart01,Si01,Lohneysen07}. In these
materials, the strong coupling limit of the on-site Kondo spin exchange
causes the Kondo screening (KS) of local moments by the conduction
electrons, yielding a Kondo paramagnetic (KP) phase. On the other hand, in
the weak coupling limit, the Kondo coupling generates an indirect
Ruderman-Kittel-Kasuya-Yosida interaction among the local moments, resulting
in either antiferromagnetism (AFM) around the half-filling of the conduction
electrons, or ferromagnetism (FM) far away from half-filling~\cite%
{Doniach77,Lacroix79}.

However, in the intermediate Kondo coupling region, the competition between
KS and magnetic correlation may produce a coexisting (CE) phase with AFM and
KS near half-filling~\cite{Zhang00,Paschen04,Friedemann09,Custers12,Isaev13}%
. The CE phase or so-called heavy fermion antiferromagnetic phase (HFAFM)
has been observed in CeCoGe$_{3-x}$Si$_{x}$~\cite{Duhwa98},and Ce$_{3}$Pd$%
_{20}$Si$_{6}$~\cite{Custers12}, etc. In Ce$_{3}$Pd$_{20}$Si$_{6}$, the
HFAFM is observed within the magnetically ordered phase, indicating the
separation of two transitions, i.e., the breakdown of Kondo screening effect and concomitant
Fermi surface reconstruction (FSR) which happens
between HFAFM and pure AFM, and the magnetic transition which occurs between
HFAFM and KP phase~\cite{Custers12}. However, studies of Hall coefficient and pressure effect
in YbRh$_{2}$Si$_{2}$ have shown that the Kondo breakdown occurs precisely
at the magnetic transition, while under Co and Ir doping, two transitions
separate~\cite{Paschen04,Friedemann09,Isaev13}. The Kondo breakdown also occurs away
from the magnetic transition in CeIn$_{3}$ and CeRh$_{1-x}$Co$_{x}$In$_{5}$~%
\cite{Harrison07,Goh08}.

To understand the novel behavior of the phase transitions in these
materials, the corresponding parameter region and the feature of the AFM
phase, CE phase, KP phase have been investigated within the framework of the
Kondo lattice model (KLM) or Kondo Heisenberg lattice model, and intensively
studied at zero temperature by mean-field approximation, variational Monte
Carlo calculations, Gutzwiller approximation, etc~\cite%
{Zhang00,Capponi01,Watanabe07,Martin08,Fabrizio08,Fabrizio13}. At
half-filling limit, the reconstruction of the energy bands leads to an
insulating state, and the ground state phases were computed with varying
Kondo coupling \cite{Zhang00,Capponi01,Watanabe07}. Away from half-filling,
the phase transitions are discussed by mean-field
approximation on the Kondo Heisenberg lattice model, and the shift from
onset to offset between the Kondo breakdown and magnetic transition is proposed to be driven by
the change of the Heisenberg exchange and the ratio of short and long-range
hopping strength~\cite{Isaev13}.

Actually, the ground state phases and their features are controlled by the
Kondo coupling $J$, the electron occupy number $n_{c}$ and the electron
hopping strength. However, the phase evolutions with these parameters have
not been fully explored yet, particularly how these phases evolute with $%
n_{c}$ and long-distance electron hoping remains an open issue~\cite%
{Lacroix79,Fabrizio13}. In order to deal with this issue, we adopt the
slave-fermion mean-field technique on the KLM, and developed a more
efficient theory to calculate the phase boundaries between various phases.
We show that the relative positions of Kondo breakdown and magnetic
transition can be shifted by both $n_{c}$ and $t^{\prime }$ on the $n_{c}$-$%
J $ plane. As $n_{c}$ decreases, two transitions get closer and then
coincide. This $n_{c}$-generating offset-to-onset structure with a triple
point in the phase diagram is related to the experimental observations of Ce$%
_{3}$Pd$_{20} $Si$_{6}$ and YbRh$_{2}$Si$_{2}$ under doping.

On the other hand, most of earlier works focused on the ground state, while
detailed theoretical studies at finite temperatures are still lacking. In
the case when the occupation number of the conduction electron $n_{c}$ is
far away from half-filling, the coexisting phase of FM and KS has
investigated at finite temperatures and the phase diagram has been derived~%
\cite{Zhang10,Liu13}. Remarkably, the boundaries separating the various
phases joint to a single point on the $J$-$T$ plane, so an interesting issue
arises as to whether the half-filled Kondo lattice system possesses similar
feature in the finite temperature phase diagram. Therefore, we devote to
study the ground state and finite-temperature phase diagram of the Kondo
lattice model at and away from half-filling, attempting to give a detailed
description of the evolution of the various phases with the Kondo coupling,
the conduction electron occupy number, the electron hopping strength and
temperature. To this end, a modified mean-field decoupling technique for the
Kondo interaction is employed, then the phase diagrams are obtained as
functions of $n_{c}$, $J$, $t^{\prime }$, and $T$ . Our method turns out to
give a compact description of the phase diagrams at both zero and finite
temperatures.

\section{Model and mean-field treatment}

We consider the spin-1/2 Kondo lattice model on a two-dimensional square
lattice with $N$ sites,
\begin{equation}
\mathcal{H}=\sum_{\mathbf{k},\sigma }(\epsilon _{\mathbf{k}}-\mu )c_{\mathbf{%
k}\sigma }^{\dag }c_{\mathbf{k}\sigma }+J\sum_{i}\mathbf{S}_{i}\cdot \mathbf{%
S}_{ic},
\end{equation}%
where $\epsilon _{\mathbf{k}}$ is the dispersion of conduction electrons,
which interact with local moments through antiferromagnetic Kondo exchange $%
J>0$, and $\mu $ denotes the chemical potential. $\mathbf{S}_{ic}=\frac{1}{2}%
\sum_{\alpha \beta }c_{i\alpha }^{\dag }\bm{\sigma}_{\alpha \beta }c_{i\beta
}$ with $\bm{\sigma}$ being the Pauli matrix, represents the spin density
for conduction electrons, while the local moments can be written in the
slave-fermion representation as $\mathbf{S}_{i}=\frac{1}{2}\sum_{\alpha
\beta }f_{i\alpha }^{\dag }\bm{\sigma}_{\alpha \beta }f_{i\beta }$, which is
subject to the restriction $\sum_{\sigma }f_{i\sigma }^{\dag }f_{i\sigma }=1$
\ imposing by a Lagrangian term $\sum_{i}\lambda _{i}(\sum_{\sigma
}f_{i\sigma }^{\dag }f_{i\sigma }-1)$. The Kondo interaction can be
decomposed into
\begin{align*}
J\sum_{i}\mathbf{S}_{i}\cdot \mathbf{S}_{ic}=& -\frac{3}{8}%
\sum_{i}(c_{i\uparrow }^{\dag }f_{i\uparrow }+c_{i\downarrow }^{\dag
}f_{i\downarrow })(f_{i\uparrow }^{\dag }c_{i\uparrow }+f_{i\downarrow
}^{\dag }c_{i\downarrow }) \\
& +\frac{1}{8}\sum_{i}(c_{i\uparrow }^{\dag }f_{i\uparrow }-c_{i\downarrow
}^{\dag }f_{i\downarrow })(f_{i\uparrow }^{\dag }c_{i\uparrow
}-f_{i\downarrow }^{\dag }c_{i\downarrow }) \\
& +\frac{1}{8}\sum_{i}(c_{i\uparrow }^{\dag }f_{i\downarrow }+c_{i\downarrow
}^{\dag }f_{i\uparrow })(f_{i\downarrow }^{\dag }c_{i\uparrow }+f_{i\uparrow
}^{\dag }c_{i\downarrow }) \\
& +\frac{1}{8}\sum_{i}(c_{i\uparrow }^{\dag }f_{i\downarrow }-c_{i\downarrow
}^{\dag }f_{i\uparrow })(f_{i\downarrow }^{\dag }c_{i\uparrow }-f_{i\uparrow
}^{\dag }c_{i\downarrow }),
\end{align*}%
where the first term represents the Kondo singlet screening effect and the
other three terms describe the triplet parings between conduction electrons
and slave fermion holes. This expression captures SU(2) invariance of the
Kondo coupling.

In order to describe the antiferromagnetism in the Kondo lattice model, two
AFM order parameters
\begin{equation}
m_{f}=\frac{1}{2}\sum_{\sigma }\sigma \langle f_{i\sigma }^{\dag }f_{i\sigma
}\rangle e^{i\mathbf{Q}\cdot \mathbf{R}_{i}},m_{c}=\frac{-1}{2}\sum_{\sigma
}\sigma \langle c_{i\sigma }^{\dag }c_{i\sigma }\rangle e^{i\mathbf{Q}\cdot
\mathbf{R}_{i}}
\end{equation}%
are introduced to decouple the longitudinal Kondo spin exchange coupling %
\cite{Zhang00}, where $\mathbf{Q}=(\pi ,\pi )$ is the AFM vector. Then the
total staggered magnetization is expressed by $M=m_{f}-m_{c}$. To
characterize the KS in the presence of AFM long-range ordering, two
different hybridization parameters on each magnetic sublattice A and B have
to be introduced \cite{Zhang00,Fabrizio13}
\begin{align*}
& V_{1}=\langle c_{iA\uparrow }^{\dag }f_{iA\uparrow }\rangle =\langle
c_{iB\downarrow }^{\dag }f_{iB\downarrow }\rangle , \\
& V_{2}=\langle c_{iB\uparrow }^{\dag }f_{iB\uparrow }\rangle =\langle
c_{iA\downarrow }^{\dag }f_{iA\downarrow }\rangle .
\end{align*}%
The conventional hybridization parameter is expressed as
\begin{eqnarray*}
V_{s} &=&\frac{1}{2}(V_{1}+V_{2}) \\
&=&\frac{1}{2}\langle c_{iA\uparrow }^{\dag }f_{iA\uparrow }+c_{iA\downarrow
}^{\dag }f_{iA\downarrow }\rangle =\frac{1}{2}\langle c_{iB\uparrow }^{\dag
}f_{iB\uparrow }+c_{iB\downarrow }^{\dag }f_{iB\downarrow }\rangle ,
\end{eqnarray*}%
while the staggered hybridization parameter is defined by
\begin{eqnarray*}
V_{t} &=&\frac{1}{2}(V_{1}-V_{2}) \\
&=&\frac{1}{2}\langle c_{iA\uparrow }^{\dag }f_{iA\uparrow }-c_{iA\downarrow
}^{\dag }f_{iA\downarrow }\rangle =\frac{-1}{2}\langle c_{iB\uparrow }^{\dag
}f_{iB\uparrow }-c_{iB\downarrow }^{\dag }f_{iB\downarrow }\rangle ,
\end{eqnarray*}%
which requires the breaking of particle-hole symmetry of the conduction
electrons, i.e., $n_{c}\neq 1$, or $t^{\prime }\neq 0$. It is seen that the
singlet channel hybridizes the $c$- and $f$-fermions with the same wave
vector, while the longitudinal exchange brings a momentum transfer $\mathbf{Q%
}$ within both $c$- and $f$-fermions, resulting in the staggered triplet
channel. The local Lagrangian constraint is replaced by a uniform one: $%
\lambda _{i}=\lambda $.

Though such a mean-field treatment, the model Hamiltonian is written in the
momentum space by the matrix form
\begin{equation}
\mathcal{H}=N\epsilon _{0}+{\sum_{\mathbf{k},\sigma }}^{\prime }\Phi _{%
\mathbf{k}\sigma }^{\dag }\mathcal{H}_{\mathbf{k}\sigma }\Phi _{\mathbf{k}%
\sigma },
\end{equation}%
where the superscript represents the summation of $\mathbf{k}$ restricted in
the magnetic Brillouin zone (MBZ) with boundaries $|k_{x}\pm k_{y}|=\pi $, a
four-component Nambu operator has been used $\Phi _{\mathbf{k}\sigma }=(c_{%
\mathbf{k}\sigma }\hspace{0.1cm}c_{\mathbf{k}+\mathbf{Q}\sigma }\hspace{0.1cm%
}f_{\mathbf{k}\sigma }\hspace{0.1cm}f_{\mathbf{k}+\mathbf{Q}\sigma })^{T}$,
and the constant term is given by $\epsilon _{0}=\frac{J}{2}%
(3V_{s}^{2}-V_{t}^{2})+Jm_{c}m_{f}-\lambda $. The Hamiltonian matrix is
given by
\begin{equation*}
\mathcal{H}_{\mathbf{k}\sigma }=\left(
\begin{array}{cccc}
\epsilon _{\mathbf{k}}-\mu & \frac{1}{2}Jm_{f}\sigma & -\frac{3}{4}JV_{s} &
\frac{1}{4}JV_{t}\sigma \\
\frac{1}{2}Jm_{f}\sigma & \epsilon _{\mathbf{k}+\mathbf{Q}}-\mu & \frac{1}{4}%
JV_{t}\sigma & -\frac{3}{4}JV_{s} \\
-\frac{3}{4}JV_{s} & \frac{1}{4}JV_{t}\sigma & \lambda & -\frac{1}{2}%
Jm_{c}\sigma \\
\frac{1}{4}JV_{t}\sigma & -\frac{3}{4}JV_{s} & -\frac{1}{2}Jm_{c}\sigma &
\lambda%
\end{array}%
\right) ,
\end{equation*}%
where $\epsilon _{\mathbf{k}}=-2t(\cos {k_{x}}+\cos {k_{y}})+4t^{\prime
}\cos {k_{x}}\cos {k_{y}}$ is the tight-binding dispersion of conduction
electrons with nearest-neighbor (NN) and next-nearest-neighbor (NNN) hoping
strength $t$ and $t^{\prime }$, respectively. In general filling case of
conduction electrons, the quasiparticle excitation spectrums can not be
derived analytically, we have to perform numerical calculations. However, at
half-filling, the particle-hole symmetry can help to simplify the related
calculations.

\section{Zero-temperature phase diagram at half-filling}

We first discuss the half-filling limit with only NN hoping $t$. In this
case, the particle-hole symmetry guarantees $\lambda =\mu =0$. Moreover, the
staggered hybridization disappears as $V_{t}=0$. Further discussions
including the influence of $n_{c}$ and NNN hoping $t^{\prime }$ will be
given in the last section. The NN hoping between conducting electrons leads
to the dispersion $\epsilon _{\mathbf{k}}=-2t(\cos {k_{x}}+\cos {k_{y}})$,
satisfying the relation $\epsilon _{\mathbf{k}+\mathbf{Q}}=-\epsilon _{%
\mathbf{k}}$. For convenience, we define $V=-\frac{3}{2}V_{s}$, then the
analytic formulas for the four dispersions are obtained by diagonalizing $%
\mathcal{H}_{\mathbf{k}\sigma }$:
\begin{equation}
\pm E_{\mathbf{k}}^{\pm }=\pm \frac{1}{\sqrt{2}}\sqrt{E_{1\mathbf{k}}\pm
\sqrt{E_{1\mathbf{k}}^{2}-E_{2\mathbf{k}}^{2}}},
\end{equation}%
where
\begin{align*}
E_{1\mathbf{k}}& =\epsilon _{\mathbf{k}}^{2}+\frac{1}{4}%
J^{2}(m_{c}^{2}+m_{f}^{2})+\frac{1}{2}J^{2}V^{2}, \\
E_{2\mathbf{k}}& =\sqrt{J^{2}m_{c}^{2}\epsilon _{\mathbf{k}}^{2}+\frac{1}{4}%
J^{4}(m_{c}m_{f}+V^{2})^{2}},
\end{align*}%
with the relation
\begin{equation*}
E_{\mathbf{k}}^{-}+E_{\mathbf{k}}^{+}=\sqrt{E_{1\mathbf{k}}+E_{2\mathbf{k}}}%
\equiv E_{\mathbf{k}}.
\end{equation*}

At zero temperature, two lower branches of spectrums $-E_{\mathbf{k}}^{\pm }$
lying below the Fermi level are full occupied, giving rise to an insulating
heavy fermion state. Two higher branches $E_{\mathbf{k}}^{\pm }$ above the
Fermi level give no contribution to the ground state energy, therefore the
ground state energy density is evaluated as
\begin{equation}
E_{g}^{CE}=[\frac{2}{3}JV^{2}+Jm_{c}m_{f}]-\frac{1}{N}{\sum_{\mathbf{k}}}E_{%
\mathbf{k}},
\end{equation}%
with the summation of $\mathbf{k}$ runs over the entire Brillouin zone. The
mean-field order parameters are determined by minimizing $E_{g}^{CE}$,
yielding the self-consistent equations
\begin{align}
& \frac{J}{4N}{\sum_{\mathbf{k}}}\frac{1}{E_{\mathbf{k}}}=\frac{m_{c}}{%
3(m_{f}-m_{c})},  \notag \\
& \frac{J^{3}}{8N}(m_{c}m_{f}+V^{2}){\sum_{\mathbf{k}}}\frac{1}{E_{\mathbf{k}%
}E_{2\mathbf{k}}}=\frac{2m_{f}-3m_{c}}{3(m_{f}-m_{c})},  \notag \\
& \frac{3J}{2N}{\sum_{\mathbf{k}}}\frac{\epsilon _{\mathbf{k}}^{2}}{E_{%
\mathbf{k}}E_{2\mathbf{k}}}=1+m_{f}/m_{c}.  \label{integrals}
\end{align}

In the whole coupling range, the pure AFM phase and KP phase should also be
examined, then the stable ground state phase corresponds to the phase with
lowest energy. When $m_{c}=m_{f}=0$, the energy density of KP phase is
obtained by
\begin{equation}
E_{g}^{KP}=\frac{2}{3}JV^{2}-\frac{1}{N}\sum_{\mathbf{k}}\sqrt{\epsilon _{%
\mathbf{k}}^{2}+J^{2}V^{2}},
\end{equation}%
where $V$ is determined by $\frac{J}{N}\sum_{\mathbf{k}}\frac{1}{\sqrt{%
\epsilon _{\mathbf{k}}^{2}+J^{2}V^{2}}}=4/3$. For the pure AFM phase with $%
V=0$, the conduction electrons and the local moments are totally decoupled,
resulting in the dispersions $E_{\mathbf{k}}^{\pm }=\pm \sqrt{\epsilon _{%
\mathbf{k}}^{2}+\frac{1}{4}J^{2}m_{f}^{2}}$ for the conduction electrons and
$E_{d}^{\pm }=\pm \frac{1}{2}Jm_{c}$ for local moments, respectively. By
performing similar self-consistent treatments, the energy for AFM phase is
found to be
\begin{equation}
E_{g}^{AF}=-\frac{1}{N}\sum_{\mathbf{k}}\sqrt{\epsilon _{\mathbf{k}%
}^{2}+J^{2}/16},
\end{equation}%
with order parameters $m_{f}=1/2$, and $m_{c}=\frac{J}{8N}\sum_{\mathbf{k}}%
\frac{1}{\sqrt{\epsilon _{\mathbf{k}}^{2}+J^{2}/16}}$.

\begin{figure}[tbp]
\hspace{-0.3cm} \includegraphics[totalheight=2.35in]{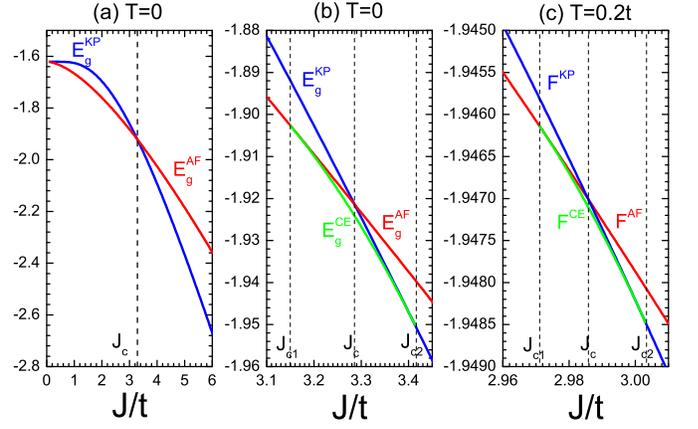}
\caption{(Color online) (a) (b) Zero-temperature energies of AFM phase $%
E^{AF}_g$, KP phase $E^{KP}_g$ and CE phase $E^{CE}_g$ vs Kondo coupling $J$%
. (c) Free energies for the three phases as functions of $J$ at temperature $%
T=0.2 t$. The CE phase exists in a narrow range between $J_{c1}$ and $J_{c2}$%
. All energies are in unit of NN hopping strength $t$. }
\label{energy}
\end{figure}

\begin{figure}[tbp]
\hspace{-0.5cm} \includegraphics[totalheight=2.4in]{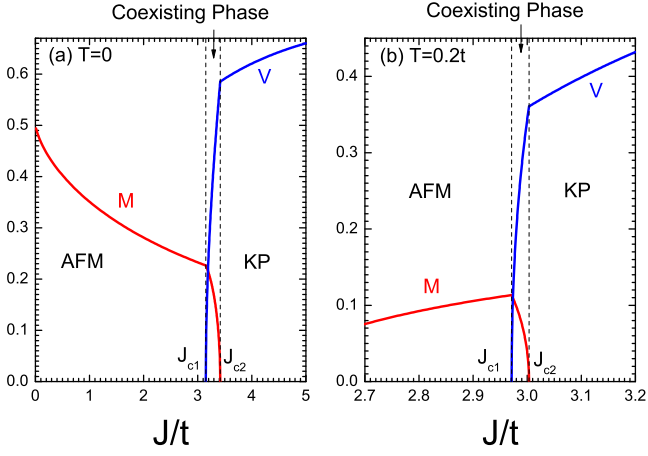}
\caption{(Color online) Staggered magnetization $M$ and Kondo hybridization $%
V$ as functions of Kondo coupling $J$ at (a) $T$=0 and (b) $T$=0.2$t$ for
the three phases. }
\label{parameters}
\end{figure}

The comparison of the ground state energies for AFM, CE and KP phases is
demonstrated in Fig. \ref{energy}(a)-(b). As expected, the competition
between ordering of local moments and formation of Kondo singlets leads to a
coexisting solution with lowest energy, indicating the stability of the CE
phase in the intermediate Kondo coupling range $J_{c1}<J<J_{c2}$~\cite%
{Zhang00}, while the pure AFM phase and the KP phase exist in the region $%
J<J_{c1}$ and $J>J_{c2}$, respectively. The derived staggered magnetization $%
M=m_{f}-m_{c}$ and KS strength $V$ are given in Fig. \ref{parameters}(a) as
a function of the Kondo coupling $J$. In the CE phase, the fluctuations of
local spins in the Kondo channel suppress the AFM order, while the staggered
magnetic order brings a rapid decrease of the KS strength. Both $M$ and $V$
vary continuously on the phase boundaries. $J_{c1}$ and $J_{c2}$ correspond
to the lower boundary of the KS state with order parameters $m_{c}\neq
0,m_{f}\neq 0,V\rightarrow 0$, and the upper boundary of the AFM state with $%
m_{c}\rightarrow 0,m_{f}\rightarrow 0,V\neq 0$, respectively. The limit $%
V\rightarrow 0$ can be replaced by setting $V=0$ in the self-consistent
equations Eq. (\ref{integrals}), because the denominators of the integral
functions are always nonzero, leading to the numerical results $%
J_{c1}=3.1498t,m_{f}=0.5,m_{c}=0.2735$. To calculate $J_{c2}$, we can simply
set $m_{c}=0,m_{f}=0$ in the integrals in Eq. (\ref{integrals}) because $E_{%
\mathbf{k}}=\sqrt{\epsilon _{\mathbf{k}}^{2}+J_{c2}^{2}V^{2}}$ and $E_{2%
\mathbf{k}}=\frac{1}{2}J_{c2}^{2}V^{2}$ are both gapped. To keep $%
m_{f}/m_{c} $ as a constant, we obtain the numerical solutions $%
J_{c2}=3.4161t,V=0.5853$.

As noticed, the energy of CE phase $E_{g}^{CE}$ is tangent to $E_{g}^{AF}$
on the edge $J_{c1}$, and to $E_{g}^{KP}$ at $J_{c2}$, respectively,
implying that the Kondo lattice system undergoes second-order phase
transitions on both phase boundaries. The coexistence of AFM and KS in the
Kondo lattice systems has been reported previously at zero temperature~\cite%
{Zhang00,Isaev13,Capponi01}. Though the proposed phase boundaries $J_{c1}$ and $J_{c2}$
are slightly different from our results, the main physical pictures remain
the same. This inconsistence comes from distinct mean-field decoupling
procedures. In these earlier studies, the Kondo exchange is decomposed
directly into longitudinal term and transversal part (which is the singlet
channel of hybridization), respectively; while in this work, the singlet and
triplet hybridization between itinerant electrons and local moments are
considered at the same level in the beginning. Therefore, our method can be
generalized straightforwardly to deal with the case away from half-filling
and with hoping beyond NN, as will be discussed in the following.

\section{Finite temperature phase diagram at half-filling}

At finite temperatures, the existence of thermal fluctuations may shift the
parameter region of the CE phase. For simplicity and without loss of
generality, we consider half-filling case $n_{c}=1$ with $t^{\prime }=0$.
Since in this situation, the mean-field Hamiltonian has been diagonalized
with the spectrums $\pm E_{\mathbf{k}}^{\pm }$, the free energy density $%
F^{CE}$ can be calculated via the partition function, leading to the result
\begin{equation}
F^{CE}=\frac{2}{3}JV^{2}+Jm_{c}m_{f}-\frac{2T}{N}\sum_{\mathbf{k},\pm }\ln
[2\cosh (E_{\mathbf{k}}^{\pm }/2T)].
\end{equation}%
It is easy to verify the equivalence of the ground state energy and above
free energy at zero-temperature limit. The mean-field parameters are
determined by minimizing $F^{CE}$, then the self-consistent equations are
derived as
\begin{align}
& \frac{4}{3J}=\frac{1}{N}\sum_{\mathbf{k},\pm }F_{\mathbf{k}}^{\pm }[1\pm
\frac{2E_{1\mathbf{k}}-J^{2}(m_{c}m_{f}+V^{2})}{2\sqrt{E_{1\mathbf{k}%
}^{2}-E_{2\mathbf{k}}^{2}}}],  \notag \\
& \frac{2m_{c}}{J}=\frac{1}{N}\sum_{\mathbf{k},\pm }F_{\mathbf{k}}^{\pm
}[m_{f}\pm \frac{2E_{1\mathbf{k}}m_{f}-J^{2}m_{c}(m_{c}m_{f}+V^{2})}{2\sqrt{%
E_{1\mathbf{k}}^{2}-E_{2\mathbf{k}}^{2}}}],  \notag \\
& \frac{1}{N}\sum_{\mathbf{k},\pm }F_{\mathbf{k}}^{\pm }[m_{c}\pm \frac{E_{1%
\mathbf{k}}m_{c}-\frac{J^{2}}{2}m_{f}(m_{c}m_{f}+V^{2})-2m_{c}\epsilon _{%
\mathbf{k}}^{2}}{\sqrt{E_{1\mathbf{k}}^{2}-E_{2\mathbf{k}}^{2}}}]  \notag \\
& =\frac{2m_{f}}{J}.  \label{self-consistent}
\end{align}%
where $F_{\mathbf{k}}^{\pm }=\frac{1}{4E_{\mathbf{k}}^{\pm }}\tanh (E_{%
\mathbf{k}}^{\pm }/2T)$. In order to draw the phase diagram on the $J$-$T$
plane, the free energies of the pure AFM phase and KP phase should also be
calculated.

The free energy of the KP phase is given by
\begin{equation}
F^{KP}=\frac{2}{3}JV^{2}-\frac{2T}{N}\sum_{\mathbf{k},\pm }\ln [2\cosh (E_{%
\mathbf{k}}^{\pm }/2T)],
\end{equation}%
with $E_{\mathbf{k}}^{\pm }=\frac{1}{\sqrt{2}}\sqrt{\epsilon _{\mathbf{k}%
}^{2}+J^{2}V^{2}/2\pm |\epsilon _{\mathbf{k}}|\sqrt{\epsilon _{\mathbf{k}%
}^{2}+J^{2}V^{2}}}$, and the equation for $V$:
\begin{equation*}
\frac{4}{3J}=\frac{1}{4N}\sum_{\mathbf{k},\pm }\frac{\tanh (E_{\mathbf{k}%
}^{\pm }/2T)}{E_{\mathbf{k}}^{\pm }}(1\pm \frac{|\epsilon _{\mathbf{k}}|}{%
\sqrt{\epsilon _{\mathbf{k}}^{2}+J^{2}V^{2}}}).
\end{equation*}%
As $T$ approaches Kondo temperature $T_{K}$, $V\rightarrow 0$, then $E_{%
\mathbf{k}}^{+}\rightarrow |\epsilon _{k}|$, $E_{\mathbf{k}}^{-}\rightarrow
0 $, therefore $T_{K}$ of the KP phase is determined by
\begin{equation}
\frac{4}{3J}=\frac{1}{2N}\sum_{\mathbf{k}}\frac{\tanh (|\epsilon _{\mathbf{k}%
}|/2T_{K})}{|\epsilon _{\mathbf{k}}|}.
\end{equation}%
For the pure AFM phase, the corresponding free energy density is written as
\begin{align}
F^{AF}& =Jm_{c}m_{f}-2T\ln [2\cosh (E_{d}^{+}/2T)]  \notag \\
& -\frac{2T}{N}\sum_{\mathbf{k}}\ln [2\cosh (E_{\mathbf{k}}^{+}/2T)],
\label{FAF}
\end{align}%
with the self-consistent equations for the AFM order parameters:
\begin{align*}
m_{c}-\frac{Jm_{f}}{N}\sum_{\mathbf{k}}\frac{\tanh (E_{\mathbf{k}}^{+}/2T)}{%
4E_{\mathbf{k}}^{+}}& =0, \\
m_{f}-\frac{1}{2}\tanh (Jm_{c}/4T)& =0.
\end{align*}%
When $T$ approaches the N\'{e}el temperature $T_{N}$, $m_{f},m_{c}%
\rightarrow 0$, then $E_{\mathbf{k}}^{+}\rightarrow |\epsilon _{\mathbf{k}%
}|,E_{d}^{+}\rightarrow 0$, thus the equation determining $T_{N}$ is derived
as
\begin{equation}
\frac{32T_{N}}{J^{2}}=\frac{1}{N}\sum_{\mathbf{k}}\frac{\tanh (|\epsilon _{%
\mathbf{k}}|/2T_{N})}{|\epsilon _{\mathbf{k}}|}.
\end{equation}

In order to derive the finite-temperature phase diagram, the free energies
of the AFM phase, the KP phase and the CE phase are compared. In Fig. \ref%
{energy}(c), the free energies of the three phases are plotted as functions
of kondo coupling strength $J$ at temperature $T=0.2t$. It can be seen that
the CE phase is stable in the coupling range $J_{c1}<J<J_{c2}$, as it
exhibits lowest free energy, and the AFM and KP phase occur in the coupling
region $J<J_{c1}$ and $J>J_{c2}$, respectively. The phase boundaries $J_{c1}$
and $J_{c2}$ now vary with temperature, and are crucial to determine the
phase diagram on the $J$-$T$ plane. Alternatively, we can consider the
characteristic temperature separating CE phase with AFM phase as a function
of $J$. On this boundary, $V$ approaches zero, so it appears as the Kondo
temperature $T_{K}^{\prime }$ in CE phase. Using Eq. (\ref{self-consistent}%
), the equations determining $T_{K}^{\prime }$ are reduced to
\begin{align}
& \frac{4}{3J}-\frac{1}{N}\sum_{\mathbf{k},\pm }\frac{\tanh (E_{\mathbf{k}%
}^{\pm }/2T_{K}^{\prime })}{4E_{\mathbf{k}}^{\pm }}[1\pm \frac{2E_{1\mathbf{k%
}}-J^{2}m_{c}m_{f}}{2\sqrt{E_{1\mathbf{k}}^{2}-E_{2\mathbf{k}}^{2}}}]=0,
\notag \\
& \frac{m_{c}}{J}-\frac{m_{f}}{N}\sum_{\mathbf{k}}\frac{\tanh (E_{\mathbf{k}%
}^{+}/2T_{K}^{\prime })}{4E_{\mathbf{k}}^{+}}=0,  \notag \\
& 2m_{f}=\tanh (Jm_{c}/4T_{K}^{\prime }),  \label{Jc1}
\end{align}%
where $E_{1\mathbf{k}}=\epsilon _{\mathbf{k}}^{2}+\frac{1}{4}%
J^{2}(m_{c}^{2}+m_{f}^{2})$, $E_{2\mathbf{k}}=Jm_{c}\sqrt{\epsilon _{\mathbf{%
k}}^{2}+\frac{1}{4}J^{2}m_{f}^{2}}$, $E_{\mathbf{k}}^{+}=\sqrt{\epsilon _{%
\mathbf{k}}^{2}+J^{2}m_{f}^{2}/4}$, and $E_{\mathbf{k}}^{-}=Jm_{c}/2$.

On the boundary between CE phase and KP phase, the AFM order vanishes, so
this phase boundary line corresponds to the N\'{e}el temperature $%
T_{N}^{\prime }$. On this edge, $m_{f}/m_{c}$ remains finite. The
self-consistent equations determining $T_{N}^{\prime }$ with varying $J$ are
simplified to
\begin{align}
& \frac{4}{3J}=\frac{1}{N}\sum_{\mathbf{k},\pm }\frac{\tanh (E_{\mathbf{k}%
}^{\pm }/2T_{N}^{\prime })}{4E_{\mathbf{k}}^{\pm }}[1\pm \frac{|\epsilon _{%
\mathbf{k}}|}{\sqrt{\epsilon _{\mathbf{k}}^{2}+J^{2}V^{2}}}],  \notag \\
& \frac{2}{J}=\frac{1}{N}\sum_{\mathbf{k},\pm }\frac{\tanh (E_{\mathbf{k}%
}^{\pm }/2T_{N}^{\prime })}{4E_{\mathbf{k}}^{\pm }}[\gamma \pm \frac{2E_{1%
\mathbf{k}}\gamma -J^{2}V^{2}}{2\sqrt{E_{1\mathbf{k}}^{2}-E_{2\mathbf{k}}^{2}%
}}],  \notag \\
& \frac{2\gamma }{J}=\frac{1}{N}\sum_{\mathbf{k},\pm }\frac{\tanh (E_{%
\mathbf{k}}^{\pm }/2T_{N}^{\prime })}{4E_{\mathbf{k}}^{\pm }}[1\pm \frac{E_{1%
\mathbf{k}}-\gamma J^{2}V^{2}/2-2\epsilon _{\mathbf{k}}^{2}}{\sqrt{E_{1%
\mathbf{k}}^{2}-E_{2\mathbf{k}}^{2}}}],  \label{Jc2}
\end{align}%
where $E_{1\mathbf{k}}=\epsilon _{\mathbf{k}}^{2}+J^{2}V^{2}/2$, $E_{2%
\mathbf{k}}=J^{2}V^{2}/2$, $E_{\mathbf{k}}^{\pm }=\frac{1}{\sqrt{2}}\sqrt{%
\epsilon _{\mathbf{k}}^{2}+J^{2}V^{2}/2\pm |\epsilon _{\mathbf{k}}|\sqrt{%
\epsilon _{\mathbf{k}}^{2}+J^{2}V^{2}}}$, and $\gamma =m_{f}/m_{c}$.

\begin{figure}[tbp]
\hspace{-0cm} \includegraphics[totalheight=2.6in]{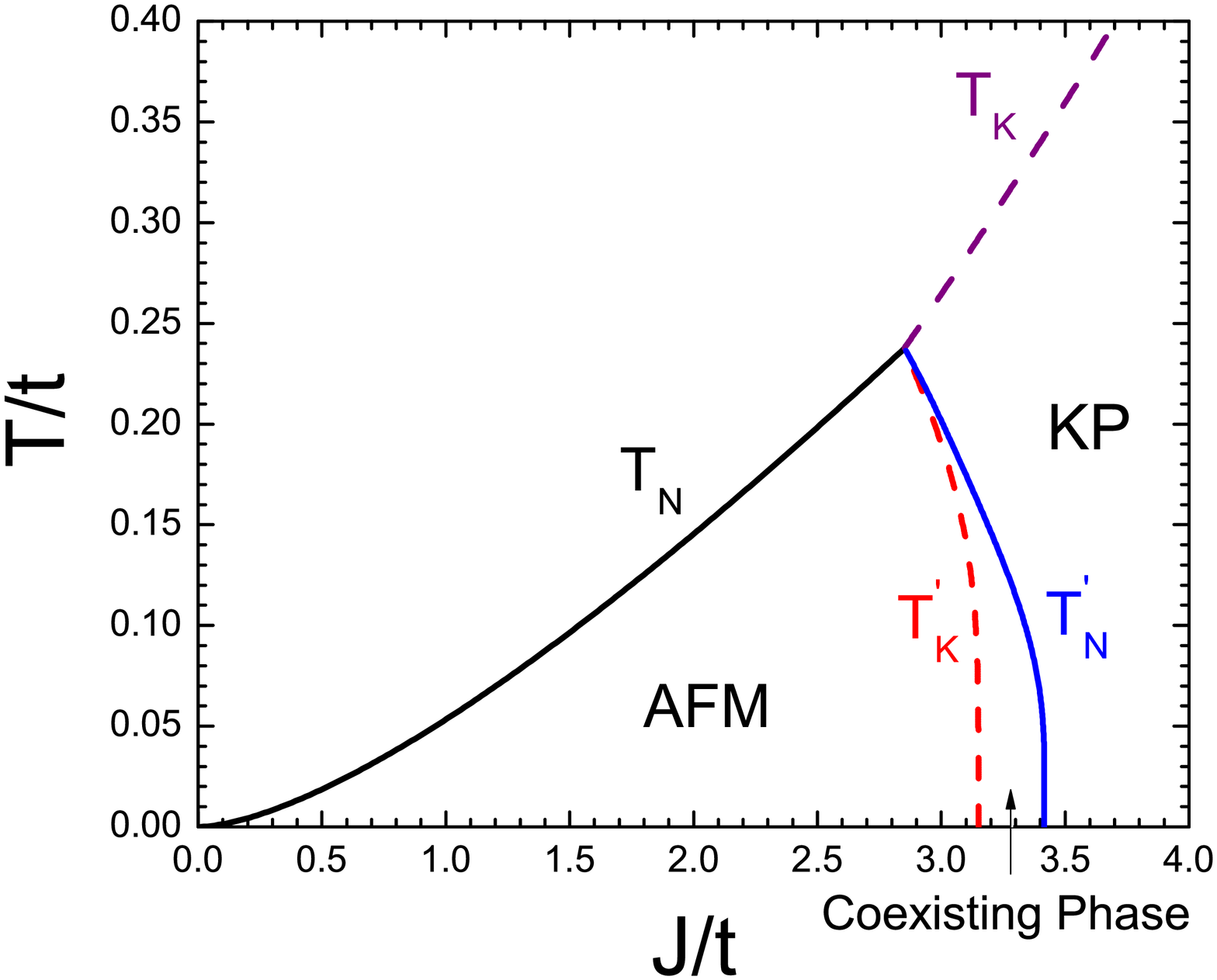}
\caption{(Color online) Finite temperature phase diagram at half-filling. In
addition to the pure AFM phase and KP phase, a coexisting phase emerges in
the area between the lines $T^\prime_K$ and $T^\prime_N$. $T_N$ and $T_K$
are N\'{e}el temperature and Kondo temperature, respectively. Three phases
converge to a single point on the $J$-$T$ plane. }
\label{phasediagram}
\end{figure}

\begin{figure}[tbp]
\hspace{-0.2cm} \includegraphics[totalheight=2.35in]{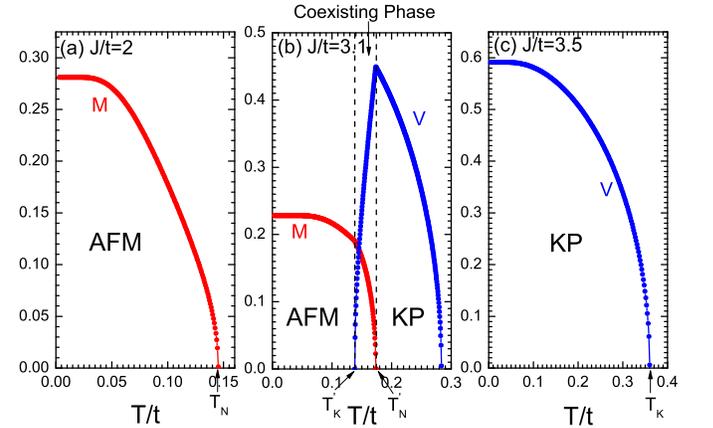}
\caption{(Color online) Temperature dependence of staggered magnetization $M$
and KS strength $V$ at fixed Kondo couplings. }
\label{parametersvsT}
\end{figure}

The critical lines $T_{K}$, $T_{N}$, $T_{K}^{\prime }$ and $T_{N}^{\prime }$
(which are all calculated as functions of $J$) necessary to determine the
finite-temperature phase diagram at half-filling case are illustrated in
Fig. \ref{phasediagram}. Both the Kondo temperature and N\'{e}el temperature
show two distinct parts. In weak Kondo coupling region, the N\'{e}el
temperature first increases from zero to a maximal value $T_{N}=0.238t$ at $%
J=2.85t$, then diminishes rapidly down to zero at a critical Kondo exchange $%
J_{c2}=3.4161t$. The reduction of $T_{N}$ in the intermediate coupling
region is due to the spin fluctuations caused by the KS in the CE phase. For
the Kondo temperature in the KP phase, it grows rapidly with increasing $J$
from $T_{K}=0.238t$ at $J=2.85t$ , while inside the CE phase, it shows a
steep reduction down to zero from $J=2.85t$ to $J_{c1}=3.1498t$. The CE
phase exists in the narrow region between $T_{K}^{\prime }$ and $%
T_{N}^{\prime }$, which contracts continuously as the temperature rises,
then finally disappears.

Notably, the four lines intersect each other at $J=2.85t,T=0.238t$, showing
that the AFM phase, the KP phase and the CE phase converge to a same point
on the $J$-$T$ plane, similar to that reported in the case of far away from
half-filling, where the ferromagnetism and KS coexist~\cite{Zhang10,Liu13}.
This feature of the phase diagram is similar to that derived by dynamic
mean-field theory~\cite{Si01}. In fact, the CE phase deduced by previous
mean-field treatments~\cite{Zhang00,Capponi01} does not converge together
with the AFM and KP phase to a single point in the phase diagram, in
contrast, the CE phase diminishes as temperature rises, and disappears
inside the AFM and KP phase. Therefore, the existence of this convergent
point deserves further verification beyond simple mean-field treatments.

The staggered magnetization $M$ and the Kondo hybridization parameter $V$
are calculated as a function of $J$ for given temperatures, illustrated in
Fig. \ref{parameters}(b), and as functions of $T$ but constant $J$,
demonstrated in Fig. \ref{parametersvsT}. On the edge between AFM phase and
CE phase (denoted by $J_{c1}$ in Fig. \ref{parameters} and by $T_{K}^{\prime
}$ in Fig. \ref{parametersvsT} (b)), $M$ varies continuously showing a kink,
then decreases and approaches zero on the upper edge $J_{c2}$ or $%
T_{N}^{\prime }$; while on the boundary between CE phase and KP phase ($%
J_{c2}$ in Fig. \ref{parameters} and $T_{N}^{\prime }$ in Fig. \ref%
{parametersvsT} (b)), the KS strength $V$ also varies continuously with a
kink and then decreases and disappears when approaching the lower edge ($%
J_{c1}$ or $T_{K}^{\prime }$). The suppression of AFM order and Kondo
hybridization by each other in CE phase is owing to the competition between
them.

\section{Ground-state phase diagram close to half-filling}

In above sections, the CE phase is studied in the half-filling case with
only NN hoping $t$ among conduction electrons, and the systems is in an
insulating phase. While away from half-filling or with NNN hoping $t^{\prime
}$ or beyond, the system no longer possesses particle-hole symmetry, thus in
addition to the singlet hybridization $V_{s}$, the triplet hybridization $%
V_{t}$ between the conduction electrons and local moments plays an important
role. Consequently, the system may possess enriched phase transitions and
phase diagram. In contrast to the mean-field methods in earlier works~\cite%
{Zhang00,Capponi01}, the optimized mean-field decoupling we employed here
can be naturally generalized to include the influence of $n_{c}$ and $%
t^{\prime }$, and various phases and phase transitions between them can be
discussed explicitly.

In general case, the quasiparticle spectrums of CE phase have to be derived
by diagonalizing $\mathcal{H}_{\mathbf{k}\sigma }$ numerically, and the
unitary transformation between the quasiparticles and $c$ and $f$ fermions
can also obtained through this computation. The six self-consistent
equations for CE phase are derived by fitting the number of $c$ and $f$
fermions to $n_{c}$ and $1$, respectively, and by the definition of
mean-filed parameters $V_{s}$, $V_{t}$, $m_{c}$, $m_{f}$. Each equation is
expressed in turns of the matrix elements of the unitary transformation.
These equations are solved iteratively until convergence is reached, then
the energy of CE phase is obtained by summing the excitations below Fermi
level. For pure AFM phase and KP phase, since the analytic spectrums exist,
these two phase can be solved by minimizing their ground-state energies. In
the AFM phase, the conduction electrons and $f$-fermions are decoupled, with
$\lambda =0$, $m_{f}=1/2$, causing a smooth dispersions $E_{d}=\pm \frac{1}{2%
}Jm_{c}$ of local spins.

In order to determine the phase boundaries among the CE phase, AFM phase and
KP phase, we have to develop an efficient theory. Considering when $%
J\rightarrow J_{c2}$, the parameters $m_{f},m_{c},V_{t}\rightarrow 0$, the
quasiparticle spectrums of CE phase can be expressed by two parts: one is
the function of $V_{s},\lambda ,\mu $, and the other can be perturbed in the
first-order of $m_{f},m_{c},V_{t}$. To do this, we rewrite the mean-field
Hamiltonian to the form
\begin{equation}
\mathcal{H}=N\epsilon _{0}+{\sum_{\mathbf{k},\sigma }}^{\prime }\Phi _{%
\mathbf{k}\sigma }^{\dag }\mathcal{H}_{\mathbf{k}\sigma }\Phi _{\mathbf{k}%
\sigma },
\end{equation}%
the operator are redefined as $\Phi _{\mathbf{k}\sigma }=(c_{\mathbf{k}%
\sigma }\hspace{0.1cm}f_{\mathbf{k}\sigma }\hspace{0.1cm}c_{\mathbf{k}+%
\mathbf{Q}\sigma }\hspace{0.1cm}f_{\mathbf{k}+\mathbf{Q}\sigma })^{T}$, and
the Hamiltonian matrix
\begin{align*}
\mathcal{H}_{\mathbf{k}\sigma }& =\left(
\begin{array}{cccc}
\epsilon _{\mathbf{k}}-\mu & -\frac{3}{4}JV_{s} & \frac{1}{2}Jm_{f}\sigma &
\frac{1}{4}JV_{t}\sigma \\
-\frac{3}{4}JV_{s} & \lambda & \frac{1}{4}JV_{t}\sigma & -\frac{1}{2}%
Jm_{c}\sigma \\
\frac{1}{2}Jm_{f}\sigma & \frac{1}{4}JV_{t}\sigma & \epsilon _{\mathbf{k}+%
\mathbf{Q}}-\mu & -\frac{3}{4}JV_{s} \\
\frac{1}{4}JV_{t}\sigma & -\frac{1}{2}Jm_{c}\sigma & -\frac{3}{4}JV_{s} &
\lambda%
\end{array}%
\right) \\
& \equiv \left(
\begin{array}{cc}
A_{\mathbf{k}} & \sigma B_{\mathbf{k}} \\
\sigma B_{\mathbf{k}} & A_{\mathbf{k}+\mathbf{Q}}%
\end{array}%
\right) .
\end{align*}

Using the Bogoliubov transformation
\begin{equation*}
\left(
\begin{array}{c}
c_{\mathbf{k}\sigma } \\
f_{\mathbf{k}\sigma }%
\end{array}%
\right) =U_{\mathbf{k}}\left(
\begin{array}{c}
\alpha _{\mathbf{k}\sigma } \\
\beta _{\mathbf{k}\sigma }%
\end{array}%
\right) =\left(
\begin{array}{cc}
u_{\mathbf{k}} & -v_{\mathbf{k}} \\
v_{\mathbf{k}} & u_{\mathbf{k}}%
\end{array}%
\right) \left(
\begin{array}{c}
\alpha _{\mathbf{k}\sigma } \\
\beta _{\mathbf{k}\sigma }%
\end{array}%
\right)
\end{equation*}%
with $u_{\mathbf{k}}^{2}+v_{\mathbf{k}}^{2}=1$, the block diagonal parts $A_{%
\mathbf{k}}$ and $A_{\mathbf{k}+\mathbf{Q}}$ are diagonalized: $U_{\mathbf{k}%
}^{+}A_{\mathbf{k}}U_{\mathbf{k}}=\mathrm{diag}(E_{\mathbf{k}+}^{(0)},E_{%
\mathbf{k}-}^{(0)})\equiv \Lambda _{\mathbf{k}}$, $U_{\mathbf{k}+\mathbf{Q}%
}^{+}A_{\mathbf{k}+\mathbf{Q}}U_{\mathbf{k}+\mathbf{Q}}=\mathrm{diag}(E_{%
\mathbf{k}+\mathbf{Q}+}^{(0)},E_{\mathbf{k}+\mathbf{Q}-}^{(0)})\equiv
\Lambda _{\mathbf{k}+\mathbf{Q}}$, where the dispersions are functions of $%
V_{s},\lambda ,\mu $, and are equal to the spectrums in the KP phase:
\begin{equation}
E_{\mathbf{k}\pm }^{(0)}=\frac{1}{2}[\epsilon _{\mathbf{k}}-\mu +\lambda \pm
\sqrt{(\epsilon _{\mathbf{k}}-\mu -\lambda )^{2}+9J^{2}V_{s}^{2}/4}].
\end{equation}%
To construct a global zero-order unitary transformation with Bogoliubov
transformation
\begin{equation*}
M_{\mathbf{k}}=\left(
\begin{array}{cc}
U_{\mathbf{k}} & 0 \\
0 & U_{\mathbf{k}+\mathbf{Q}}%
\end{array}%
\right) ,
\end{equation*}%
which acts on the Hamiltonian matrix leads to%
\begin{equation*}
M_{\mathbf{k}}^{T}\mathcal{H}_{\mathbf{k}\sigma }M_{\mathbf{k}}=\left(
\begin{array}{cc}
\Lambda _{\mathbf{k}} & \sigma D_{\mathbf{k}} \\
\sigma D_{\mathbf{k}}^{T} & \Lambda _{\mathbf{k}+\mathbf{Q}}%
\end{array}%
\right) .
\end{equation*}%
The off-diagonal elements hybrid the zero-order eigenstates with each other,
hence bring a correction to the dispersions: $E_{\mathbf{k}\pm }=E_{\mathbf{k%
}\pm }^{(0)}(V_{s},\lambda ,\mu )+E_{\mathbf{k}\pm
}^{(1)}(m_{c},m_{f},V_{t}) $, where $E_{\mathbf{k}\pm }^{(1)}$ are easily
obtained by these off-diagonal elements using perturbation theory. To the second order of $(m_{c},m_{f},V_{t})$, the
ground-state energy density near $J_{c2}$ can be divided into two parts $%
E_{g}^{CE}=E_{g}^{(0)}(V_{s},\lambda ,\mu )+E_{g}^{(1)}(m_{c},m_{f},V_{t})$,
where
\begin{align}
E_{g}^{(0)}& =\frac{2}{N}\sum_{\mathbf{k},\pm }\theta (-E_{\mathbf{k}\pm
}^{(0)})E_{\mathbf{k}\pm }^{(0)}+\frac{3}{2}JV_{s}^{2}-\lambda +\mu n_{c},
\notag \\
E_{g}^{(1)}& =\frac{2}{N}\sum_{\mathbf{k},\pm }\theta (-E_{\mathbf{k}\pm
}^{(0)})E_{\mathbf{k}\pm }^{(1)}-\frac{1}{2}JV_{t}^{2}+Jm_{c}m_{f}.
\end{align}%
Minimization of $E_{g}^{(0)}$ with respect to $V_{s},\lambda ,\mu $ gives
rise to three self-consistent equations, while differentiating $E_{g}^{(1)}$
with $(m_{c},m_{f},V_{t})$ produces other three equations. At $J_{c2}$, $%
(m_{c},m_{f},V_{t})\rightarrow 0$, but $m_{f}/m_{c}$ and $V_{t}/m_{c}$
remain finite. Solving the six equations, $J_{c2}$ and the value of $V_{s}$,
$\lambda $, $\mu $, $m_{f}/m_{c}$, $V_{t}/m_{c}$ on this boundary are
calculated.

\begin{figure}[tbp]
\hspace{-0.4cm} \includegraphics[totalheight=2.35in]{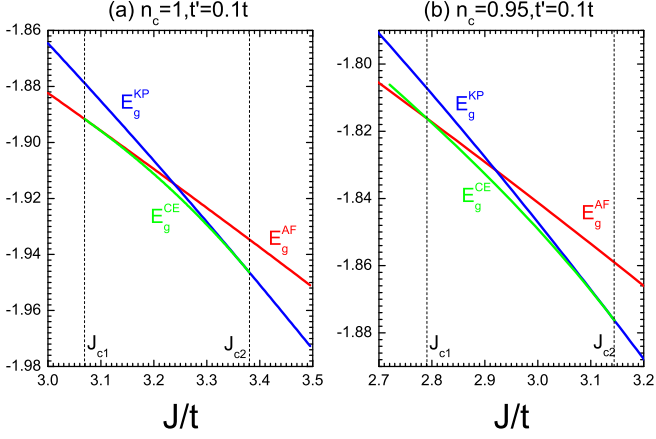}
\caption{(Color online) Energies of AFM phase $E_{g}^{AF}$, KP phase $%
E_{g}^{KP}$ and CE phase $E_{g}^{CE}$ vs Kondo exchange $J$. All energies
are in unit of NN hopping $t$. }
\label{energy2}
\end{figure}
\begin{figure}[tbp]
\hspace{-0.3cm} \includegraphics[totalheight=2.7in]{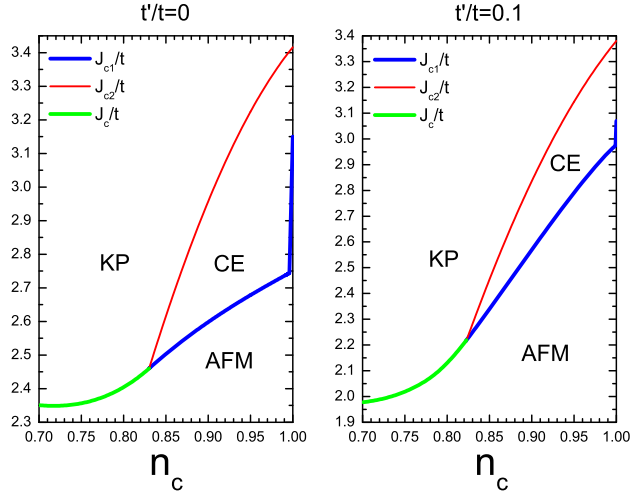}
\caption{(Color online) Ground-state phase diagram of KLM near half-filling.
As $n_{c}$ reduces, the collapse of KS at $J_{c1}$ and the magnetic
transition at $J_{c2}$ get closer and finally converge. Thick and thin lines
denote first and second order phase transitions, respectively. }
\label{phasediagram2}
\end{figure}

\begin{figure}[tbp]
\hspace{-0.4cm} \includegraphics[totalheight=2.35in]{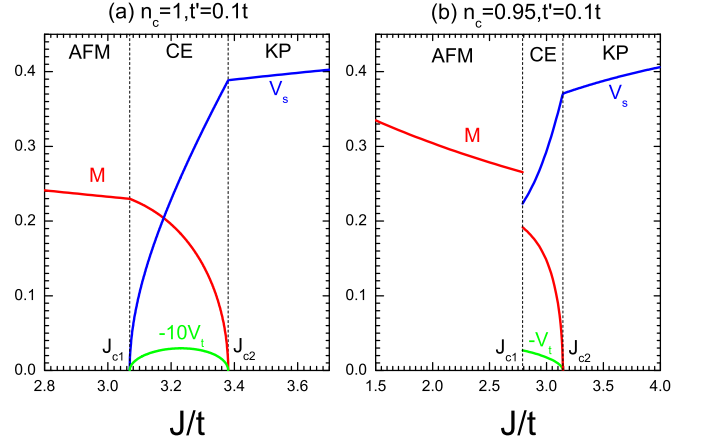}
\caption{(Color online) Staggered magnetization $M$, Kondo hybridization $%
V_{s}$ and triplet hybridization $V_{t}$ as functions of Kondo coupling $J$.
}
\label{parameters2}
\end{figure}

\begin{figure}[tbp]
\centering
\includegraphics[totalheight=2.5in]{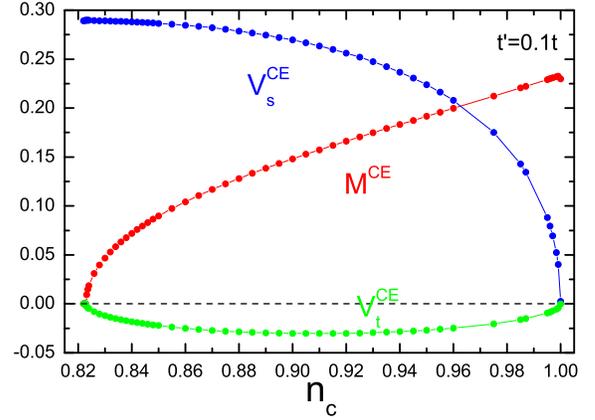}
\caption{(Color online) Staggered magnetization $M$, Kondo hybridization $%
V_{s}$ and triplet hybridization $V_{t}$ on $J_{c1}$ as functions of $n_{c}$
for CE phase. At $n_{c}=1$, $V_{s}$ and $V_{t}$ approach zero (see Fig. \ref%
{parameters2}a), while on the triple point ($n_{c}^{\ast }$=0.8228 for $%
t^{\prime }/t=0.1$), $M$ and $V_{t}$ disappear. }
\label{parasonJc1}
\end{figure}

In Fig. \ref{energy2}, the energies of the pure AFM phase, the CE phase and
the KP phase are plotted with varying $J$, and the derived critical Kondo
coupling $J_{c1}$ between AFM and CE phase on which the two phases have
equal energy has been given as a function of $n_{c}$ and $t^{\prime }$ in
Fig. \ref{phasediagram2}. At half-filling, the pure AFM phase is separated
with the CE phase by a second-order phase transition at $J_{c1}$, and on
this boundary, $M$ varies continuously, while $V_{s}$ and $V_{t}$ approach
zero, see Figs. \ref{parameters2}(a). For $n_{c}<1$, the transition at $%
J_{c1}$ changes to a first-order one, as indicated by the kink in $E_{g}$
(Fig. \ref{energy2}(b)) and the discontinuity of $M$, $V_{s}$ and $V_{t}$ at
$J_{c1}$ (Fig. \ref{parameters2}(b)). Note that $V_{s}$ and $V_{t}$ remain
finite at $J_{c1}$ for $n_{c}<1$. At $J_{c2}$, a second-order phase
transition between CE and KP phase takes place, as seen by the tangency of
ground state energy at $J_{c2}$ in Fig. \ref{energy2}. $E_{g}$, $V_{s}$, $%
V_{t}$, $M$ all vary continuously with $J$ at $J_{c2}$, at which $V_{t}$, $M$
approach zero. Moreover, the NNN hoping $t^{\prime }$ can shift both
boundaries. We find a sudden jump of $J_{c1}$ at $n_{c}=1$ (see Fig. \ref%
{phasediagram2}), this feature can be understood by the discontinuity of $M$
and $E_{g}^{CE}$ vs $n_{c}$ on $J_{c1}$ ( see Fig. \ref{parasonJc1}) and may
attribute to the change of topology, i.e., from the first order transition
at $n_{c}<1$ to second-order transition at $n_{c}=1$. Only $J_{c2}$
represents a real phase transition, because at this boundary the staggered
magnetization $M$ vanishes from AFM to KP phase.

\begin{figure}[tbp]
\hspace{-0.3cm} \includegraphics[totalheight=2.9in]{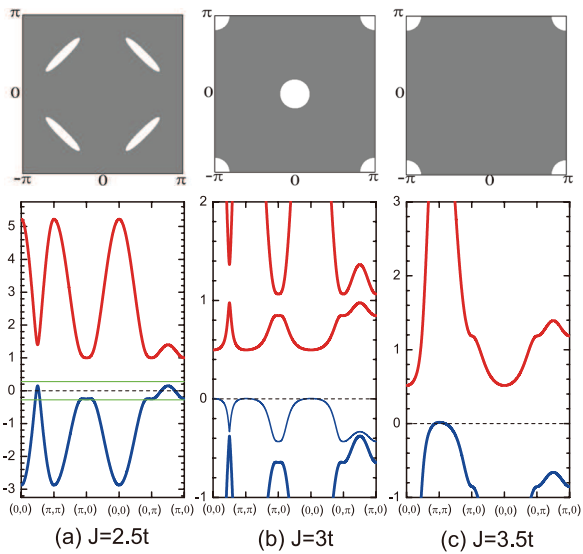}
\caption{(Color online) Spectrums and Fermi surfaces of (a) AFM phase, (b)
CE phase and (c) KP phase at $n_c=0.95$ and $t^{\prime}/t=0.1$. The shaded
areas are the occupied Fermi sea, the white zones denote Fermi holes. }
\label{Fermisurface}
\end{figure}

The spectrums and Fermi surface topological structures of these three phases
are shown in Fig. \ref{Fermisurface}. The itinerant electrons and local
moments are completely decoupled in pure AFM phase, leading to a hole-like
Fermi surface around $(\pi /2,\pi /2,)$ at $t^{\prime }=0.1$, which is
consisted of the conduction electrons only and the Luttinger volume equals $%
n_{c}S_{F}$ , where $S_{F}$ is the area of Brillouin zone; while in the CE
phase, the hybridization of conduction electrons and local moments
constructs a hole-like Fermi surface around $(0,0)$ and $(\pi ,\pi )$
points, with Fermi surface volume $n_{c}S_{F}$. The change of Fermi surface
topology between AFM and CE phase indicates a first-order Lifshitz
transition at $J_{c1}$. For the KP phase, a hole-like Fermi surface exists
around $(\pi ,\pi )$ with Fermi surface volume $\frac{1+n_{c}}{2}S_{F}$
containing both $c$- and $f$-fermions.

We have calculated $J_{c1}$ and $J_{c2}$ as $n_{c}$ varies from $0.7$ to $1$%
. Notably, as $n_{c}$ decreases, $J_{c1}$ and $J_{c2}$ approach each other
and finally intersect at $n_{c}^{\ast }=0.8228,J^{\ast }=2.2181t$ for $%
t^{\prime }/t=0.1$, and at $n_{c}^{\ast }=0.8301,J^{\ast }=2.4604t$ for $%
t^{\prime }=0$, respectively. This result indicates that the CE phase region reduces with
decreasing $n_{c}$ and finally disappears at a triple point. As $n_{c}$
decreases further, the AFM and KP phases are separated by a first-order
transition at $J_{c}$, at which the AFM and KP phase shares equal energy.
The ground state phase diagram of the KLM are summarized in Fig. \ref%
{phasediagram2} on the $n_{c}$-$J$ plane for given hopping parameter $%
t^{\prime }/t=0$ and $0.1$. When $n_{c}>n_{c}^{\ast }$, the collapse of $%
V_{s}$ at $J_{c1}$ and the magnetic transition occurring at $J_{c2}$
separate. While $n_{c}<n_{c}^{\ast }$, the KS collapses precisely at the
magnetic transition point. Similar phase diagrams are also given by mean-field treatments,
Gutzwiller approximation and variational Monte Carlo approach~\cite{Isaev13,Fabrizio08,Watanabe07}.
This offset-to-onset transition between Kondo breakdown and magnetic transition as $n_{c}$ decreases may account for
the experimental observations for CeIn$_{3}$ and CeRh$_{1-x}$Co$_{x}$In$_{5}$%
~\cite{Harrison07,Goh08}, and YbRh$_{2}$Si$_{2}$ under Co and Ir doping and
external pressure~\cite{Paschen04,Friedemann09}.

The triple point ($n_{c}^{\ast },J^{\ast }$) in our phase diagram can be
shifted by $t^{\prime }/t$, so in our mechanism this offset-to-onset
transition can also be generated by varying $t^{\prime }/t$, similar to that
proposed in Ref. \cite{Isaev13}. Chemical or external pressure may
simultaneously change $t^{\prime }/t$ and $n_{c}$, so which path cut in our
phase diagram corresponding to these experiments is not clear. Experimental
studies of the existence of $n_{c}^{\ast }$ may be particularly interesting.
The KP phase possesses larger Fermi surface than AFM phases, consequently,
the transition from AFM to KP at $J_{c}$ when $n_{c}<n_{c}^{\ast }$ may
induce a abrupt change of Hall coefficient as in YbRh$_{2}$Si$_{2}$, where
the FSR was observed via Hall effect at the onset
of magnetic transition~\cite{Friedemann09}.

\section{Conclusion}

In summary, we have performed an optimized mean-field decoupling of the
Kondo lattice model near half-filling at both zero and finite temperatures.
In addition to the pure AFM phase in weak Kondo coupling range and the Kondo
paramagnetic phase in relatively strong coupling range, a distinct phase
coexisting AFM order with Kondo hybridization arises in the intermediate
Kondo exchange region, and the ground state phase diagram has been
determined as function of the Kondo coupling, electron concentration and
electron hopping. In particular, for the coexisting phase, we found a finite
staggered triplet hybridization between local moments and conduction
electrons. We also develop an efficient method to calculate the phase
boundaries. The characteristic parameters and Fermi surface structures of
these phases and the phase transitions between them have been discussed
explicitly. We have further found a mechanism explaining the offset-to-onset
transition between Kondo breakdown and magnetic transition, which is driven
by the decreasing of electron number $n_{c}$. This mechanism may account for
the separation of the two transitions in YbRh$_{2}$Si$%
_{2}$ under doping, and the existence of triple point ($n_{c}^{\ast
},J^{\ast }$) in the phase diagram deserves deep experimental investigation.
At half-filling limit, a finite-temperature phase diagram has been
determined on $J$-$T$ plane. As temperature rises, the region of this
coexisting phase diminishes continuously then finally converges to a single
point, together with the pure AFM phase and KP phase, which may require
further theoretical and experimental verification.

\acknowledgments

H. Li acknowledges the support by Scientific Research Foundation of Guilin
University of technology. Y. Liu thanks Yifeng Yang for stimulating discussions
and acknowledges the support by China Postdoctoral Science Foundation.
G. M. Zhang acknowledges the support of NSF-China through Grant No. 20121302227.

\end{document}